\documentclass[letterpaper, 10 pt, journal, twoside]{IEEEtran}

\usepackage{graphics} 
\usepackage{epsfig} 
\usepackage{amsmath} 
\usepackage{amssymb}  
\usepackage{algorithm}
\usepackage{mathrsfs}
\usepackage{mathtools}
\usepackage{multirow}
\usepackage{comment}
\usepackage{color}
\usepackage{cite}
\usepackage{url}
\usepackage{soul}
\usepackage{listings}
\usepackage[utf8]{inputenc}
\usepackage[T1]{fontenc}
\usepackage[english]{babel}
\usepackage{xcolor}
\usepackage[normalem]{ulem}

\usepackage{xpatch} 
\usepackage{subcaption}

\newtheorem{rem}{Remark}

\usepackage{graphicx}
\usepackage{graphics}
\usepackage{amssymb}
\usepackage{amsmath}
\usepackage{color}
\usepackage{xspace}
\usepackage{algpseudocode}
\usepackage{bbm}
\usepackage{comment}
\usepackage{algorithm}
\usepackage{algorithmicx}
\usepackage{soul}
\usepackage{array}
\usepackage[bottom]{footmisc}
\usepackage{tikz}
\usetikzlibrary{shapes,arrows}
\usetikzlibrary{decorations.pathreplacing,decorations.markings,decorations.pathmorphing,decorations.shapes}
\usetikzlibrary{positioning,fit}
\usetikzlibrary{calc}
\usepackage{multirow}
\usepackage{cancel}
\usepackage{hyperref} 
\usepackage{xr}

\tikzstyle{block}=[draw opacity=0.7,line width=1.4cm]

\definecolor{CranJ}{cmyk}{0,0.69,0.54,0.04} 
\definecolor{PinkJ}{cmyk}{0,0.71,0.43,0.12} 
\definecolor{Cran}{cmyk}{0,0.73,0.41,0.29} 
\definecolor{VRed}{cmyk}{0,0.75,0.25,0.2} 
\definecolor{ORed}{cmyk}{0,0.75,0.75,0} 
\definecolor{CBlue}{cmyk}{1,0.25,0,0} 
\setstcolor{VRed}
\usepackage{bm}

\usepackage{caption}

\newlength\myindent


\tikzset{cloud/.pic={
\node[cloud, cloud puffs=10.8,cloud puff arc=110, aspect=2, draw, text width=3cm
    ] () at (0,0) {\tikzpictext};
}}

\newcommand{\VV}{\mathcal{V}}

\newcommand{\real}{{\mathbb{R}}}
\newcommand{\reals}{{\mathbb{R}}}

\newcommand{\Lnorm}{\left\|}
\newcommand{\Rnorm}{\right\|}

\newcommand{\vect}[1]{\boldsymbol{\mathbf{#1}}}

\newcommand{\kronecker}{\raisebox{1pt}{\ensuremath{\:\otimes\:}}}

 \newcommand{\boxend}{\hfill \ensuremath{\Box}}
\newcommand{\oprocendsymbol}{\hbox{$\bullet$}}
\newcommand{\oprocend}{\relax\ifmmode\else\unskip\hfill\fi\oprocendsymbol}

\allowdisplaybreaks

\newtheorem{assump}{Assumption}

\makeatletter
\renewcommand*{\@opargbegintheorem}[3]{\trivlist
      \item[\hskip \labelsep{\emph{ #1\ #2}}] \emph{(#3):}\ \itshape}
\makeatother

\usepackage{tcolorbox}
\definecolor{mycolor}{rgb}{0.122, 0.435, 0.698}
\makeatletter
\newcommand{\mybox}[1]{%
  \setbox0=\hbox{#1}%
  \setlength{\@tempdima}{\dimexpr\wd0+13pt}%
  \begin{tcolorbox}[colframe=mycolor,boxrule=0.5pt,arc=4pt,
      left=6pt,right=6pt,top=6pt,bottom=6pt,boxsep=0pt,width=\@tempdima]
    #1
  \end{tcolorbox}
}
\makeatother

\usepackage{cite}

\makeatletter
\def\hlinewd#1{%
\noalign{\ifnum0=`}\fi\hrule \@height #1 \futurelet
\reserved@a\@xhline}
\makeatother

\begin{document}

\title{Stein Coverage: a Variational Inference Approach to Distribution-matching Multisensor Deployment 
}

\author{Donipolo Ghimire~~and~~ Solmaz S. Kia, \emph{Senior Member, IEEE}
\thanks{The authors are with the Department of Mechanical and Aerospace Engineering, University of California, Irvine, CA 92697 {\tt \{dghimire,solmaz\}@uci.edu}. This work was supported by NSF award ECCS 1653838.}
}


\maketitle

\begin{abstract}

This paper examines the spatial coverage optimization problem for multiple sensors in a known convex environment, where the coverage service of each sensor is heterogeneous and anisotropic. 
We introduce the Stein Coverage algorithm, a distribution-matching coverage approach that aims to place sensors at positions and orientations such that their collective coverage distribution is as close as possible to the event distribution. To select the most important representative points from the coverage event distribution, Stein Coverage utilizes the Stein Variational Gradient Descent (SVGD), a deterministic sampling method from the variational inference literature. An innovation in our work is the introduction of a repulsive force between the samples in the SVGD algorithm to spread the samples and avoid footprint overlap for the deployed sensors. After pinpointing the points of interest for deployment, Stein Coverage solves the multisensor assignment problem using a bipartite optimal matching process. Simulations demonstrate the advantages of the Stein Coverage method compared to conventional Voronoi partitioning multisensor deployment methods.

\end{abstract}

\section{Introduction}
\begin{figure}[!t]
    \centering
    \vspace{0.2in}
     \includegraphics[width=0.5\textwidth]{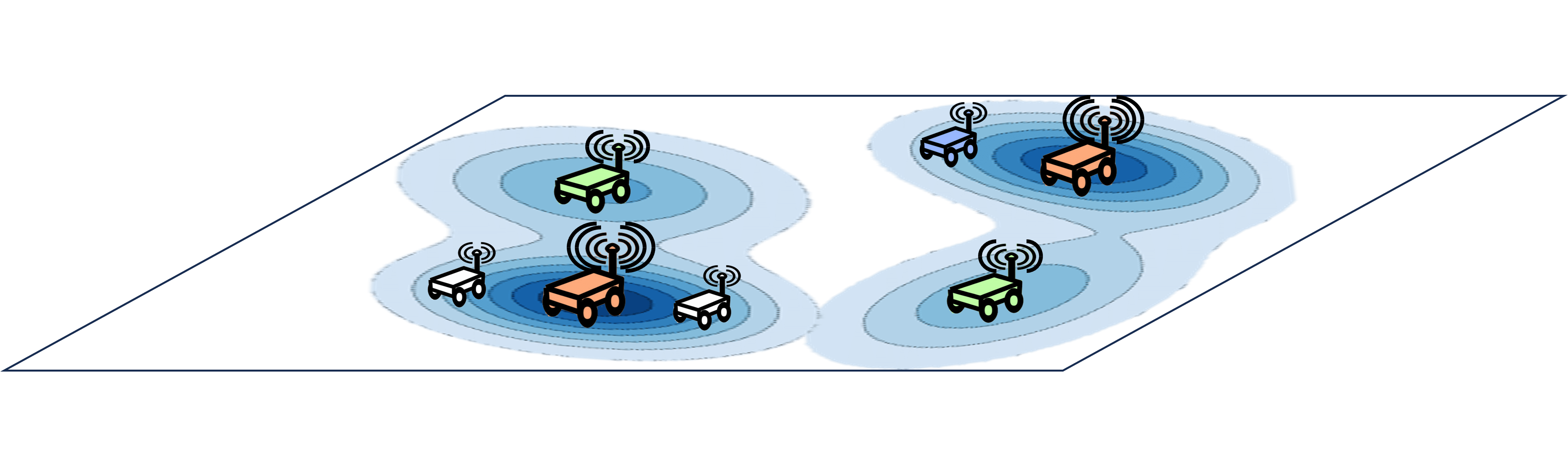} 
    \caption{{\small An example of a multisensor deployment with a limited number of sensors available. The sensors are heterogenous represented by the different sizes and colors. The aim is to place the sensors in positions that will yield the highest benefit, that is, providing service to more densely populated areas.}}
    \label{fig::prob_form} 
\end{figure}

In recent years, the decrease in the cost of embedded microsensing and wireless communication technologies has led to increased use of wireless sensor networks for tasks such as surveillance, data collection, situational awareness, improved wireless coverage, and environmental monitoring. However, the operational cost is still prohibitive in deploying an unlimited number of sensors to achieve full coverage. The coverage problem involves strategically placing a limited number of sensors across the area of interest in such a way that the spatial distribution of the sensors can capture all the essential environmental features. This paper examines a multisensor deployment problem for coverage over an event distribution in a known convex environment, where the coverage service of each sensor is heterogeneous and anisotropic, see Fig.~\ref{fig::prob_form}. The aim is to have the final configuration of the sensors achieve a topological distribution that is similar to the spatial distribution of the events induced by the targets on the ground.

\textit{Literature Review}: This research focuses on a problem in the field of sensor deployment for coverage~\cite{OA-19,NG-ES-ME:22,AF-RZ-AP:20,JC-SM-FB:04,AL-GJ-FS:18}. In coverage problems with a limited number of sensors, the essence of the proposed solutions is to divide the area into subregions and assign a sensor to each one, with the goal of maximizing a certain coverage measure.
This problem is related to many other problems, such as various resource allocation problems and the facility location problem, and is considered NP-Hard ~\cite{NM-KS:84, OA-19}. 

Among the proposed multisensor coverage strategies, geometric approaches like Voronoi diagrams, which use expected sensing cost as the Euclidean distance between sensor position and event location~\cite{MC-DR:09, JC-SM-FB:04} uniquely combines both deployment and allocation in a distributed manner using gradient descent. This approach decouples the problem of optimal sensor allocation from the problem of sensor configuration, as each sensor only needs to consider its own Voronoi cell and its local neighbors. 
Beyond partitioning, the Voronoi-based approach can consider two additional components, such as the sensing model of an agent and the map over each point in the coverage region, known as the event density function. A Voronoi-based approach can be used to model the heterogeneity of sensors using the multiplicative weighted version of Voronoi diagrams(power diagrams)~\cite{OA-DK:16,AK-SM:08, MS-ME:18}. On the other hand, the density functions can be used 
to bias the importance of the areas for coverage. Although these Voronoi-based approaches have nice properties such as convexity and dual triangulability, power diagrams may have empty cells associated with, and the locally optimal solutions provided by these methods strongly depend on proper initialization~\cite {AK-AS-GC:2008, AL-NV-JV:03}. Another downside of Voronoi partitioning methods is that the cells created by partitioning extend beyond the sensors' footprint, implicitly considering a coverage larger than the immediate sensing range. Therefore, the sensors do not necessarily get assigned to areas of importance that fall in the actual sensor footprint of the sensors.

In recent years, an alternative deployment approach has been proposed in the literature that aims to match the collective quality of service (QoS) of the sensors with the density distribution of the coverage area~\cite{OA-19,DI-YI-HY:21,VK-SM:22, YC-SK:22, DG-KS:23}.  This approach considers a more realistic model for the QoS of the sensors as a spatial distribution. 
For example~\cite{OA-19} proposed a soft and hard sensor allocation framework using a statistical coverage quality measure based on distribution matching using f-divergence between the event detection probability and spatial sensing model of sensors. Alternatively,~\cite{YC-SK:22} and \cite{DG-KS:23} proposed to cluster the area of interest using techniques such as Gaussian mixture model (GMM) and K-means clustering, respectively, to identify the potential deployment points based on the spatial distribution of the coverage subjects. Then, cast the sensor deployment problem as an optimal assignment problem. 

The deployment algorithm proposed in this paper falls into the category of distribution-matching multisensor deployment methods. We present a novel approach to determining the most suitable locations for sensor deployment by utilizing a variational inference technique. Variational inference allows drawing `super samples' from the density distribution of the coverage subjects. These samples in the statistical sense are the closest representatives of the density distribution, thus capturing the most important points in the coverage area.
In particular, we use the Stein Variational Gradient Descent (SVGD) method~\cite{QL-JL-MJ:16} to draw the `super samples'. Compared with other sampling methods such as Monte Carlo methods, SVGD is known to achieve good approximation even with a very small number of samples (particles).  SVGD is a deterministic gradient-based sampling algorithm for approximate inference. Given a probability density function $p(x)$, SVGD finds a set of samples (particles) to approximate $p(x)$ by a simple iterative update that resembles a steepest descent gradient method; for more see Section~\ref{sec::ML_opt}. However, our approach is not the immediate application of the SVGD algorithm. The sampling methods are oblivious to the coverage footprint of the sensors and can result in finding the candidate deployment points that will result in significant overlap in the coverage provided by the sensors as shown in Fig.~\ref{fig:stein_vari_algo}. The overlapping multisensor coverage is not only a waste of coverage resources but also can result in undesirable interference between the sensors' services as in wireless coverage problems. To consider the effective coverage footprint of the sensors, we include a repulsive force between the samples in the SVGD algorithm to spread the samples to avoid overlap for the deployed sensors.   The result is a pruned region of space, where the set of points determines the regions of interest and helps to create a utility map of the environment with the event distribution. This approach significantly reduces our search space for multisensor deployment. Once we generate the low-cost and constraint-satisfying utility map, we frame our problem as an assignment problem and solve the problem using the Hungarian Algorithm. We refer to our proposed deployment algorithm as {\sf{Stein Coverage}}.
We demonstrate the performance of our algorithm for heterogeneous sensors with both isotropic and anisotropic footprints.

\emph{Notation}: We let $\reals$, $\mathbb{Z}$, and $\mathbb{N}$
denote the set of reals, integers, and natural numbers, respectively. The positive and nonnegative subspaces of these number spaces are indicated by indexes $>0$ or $\geq0$, respectively. For any closed and compact set $\mathcal{C}\subset \real^2$, $\partial \mathcal{C}$ is its boundary. 

\section{Problem Definition}
\label{sec::prob}
We are considering a time-invariant event distribution function that is known a priori, $p(x): \mathcal{W} \rightarrow \mathbb{R}_{\geq 0}$, which represents the probability that some event takes place in a two-dimensional space $\mathcal{W} \subset\real^2$. The probability density represents a finite set of $m$ targets that are densely scattered over the planar space $\mathcal{W}$. These targets can be real such as humans or animals, information sources, pollution spills, forest fires, etc. Suppose $\mathcal{A}=\{1,\cdots,m\}\subset\mathbb{N}$, represents the set of $m$ sensors that can be identical (homogeneous) or non-identical (heterogeneous).
Every sensor $i\in\mathcal{A}$ provides a coverage service, e.g. wireless coverage, data harvesting, or event detection. The quality of service of each sensor is described by a spatial coverage distribution $s_i(x|x_i, \theta_i)$ where $x_i$ is the position and $\theta_i$ is the sensor orientation. The support for this distribution can be $\mathbb{R}^2$, but we assume that there is an effective coverage footprint $\mathcal{C}_i(x|x_i,\theta_i)\subset\mathbb{R}^2$ around any point $x_i$ at which the sensor is placed.  $\mathcal{C}_i(x|x_i,\theta_i)$ and $s_i(x|x_i, \theta_i)$ can be considered as rigid bodies. 
\begin{assump}\label{assump::cover_R}
    There exists $\mathsf{R}\in\mathbb{R}_{>0}$ such that $\|x-x_i\|\leq \mathsf{R}$ for any $x\in\partial\mathcal{C}_i(x|x_i,\theta_i)$
    for all $i\in\mathcal{A}$.
\end{assump}

In the statistical sense, the ultimate goal for optimal coverage is finding $(x_i,\theta_i)$ configuration for each sensor $i\in\mathcal{A}$ such that the collective distribution of $\frac{1}{m}\sum_{i\in\mathcal{A}}s_i(x|x_i,\theta_i)$ of the team is as similar as possible to the distribution of $p(x)$. The similarity between the distributions can be measured using various statistical measures, e.g., Kullback-Leibler (KL) divergence~\cite{SA-AC:10}. However, this optimization problem is highly nonlinear and computationally challenging. Therefore, the literature resorts to solving the problem under various assumptions, such as an isotropic or uniform coverage distribution for sensors or seeking suboptimal solutions.


When the sensors are homogeneous and their coverage distribution is uniform over their effective coverage range, one of the well-known frameworks to solve the multisensor coverage problem is to assign each sensor to a weighted centroid point of Voronoi partition $ \mathcal{V}(.)=\{\mathcal{V}_{1}, \mathcal{V}_{2},\cdots\mathcal{V}_N \}$ or power diagrams $ \mathcal{P}(.)=\{\mathcal{P}_{1},\mathcal{P}_{2},\cdots, \mathcal{P}_{N}\}$ obtained by using a continuous-time version of Lloyd's algorithm. This method drives the weighted centroid points to observe events distributed according to $p(x)$ by minimizing a locational optimization cost function based on a Euclidean metric. This continuous optimization can be very costly when the domain is very large and the decision variables are high dimensional. There is also no guarantee of convergence to the global minimum of the coverage merit function~\cite{OA-19,OA-DK:16}. 

Our goal is to deploy the sensors in a distribution-matching fashion at points in the coverage area where their collective service distribution best matches the distribution of the coverage subjects.
To do this, we aim to obtain a set of discrete samples 
from the most informative regions using a functional gradient method while accounting for overlapping coverage between sensors.  SVGD, Kernel herding, and Markov Chain Monte Carlo (MCMC) are some techniques that can be employed to facilitate the important sample selection process. Our method uses the Kernelized Stein discrepancy, a new discrepancy statistic for measuring differences between two probability distributions, that enables an inference problem, which looks for 
\begin{align}
    \vspace{-2.5 in}
  \min_{q}\, \mathcal{KL}(q(x) \,\| \,p(x))  \quad s.t. \quad q(g(x) \geq 0) = 1  
\end{align} where $q$ is the empirical distribution of a set of points generated from an arbitrary (Gaussian, Uniform, Beta, etc) distribution, and $g(x)$ is the soft constraint imposed on the sensor locations. Our method is a spatially regulated SVGD that introduces a metric that encourages the generated sensor locations to maintain a desired minimum distance between each other. This ensures that the sensors are strategically placed and collectively cover the dense and important regions of the environment. Similarly, for heterogeneous sensors, these final sensor configurations resolve an assignment problem, where the assignment cost for each sensor is determined by the
dissimilarity of the coverage utility and the distribution of the targets near the potential assignment location.


\section{A brief overview of Stein Variational Gradient Descent Preliminary material}\label{sec::ML_opt}

In this section, for the convenience of the reader and also to introduce the notation and definitions that we use in developing our algorithm, we give a brief overview of the Stein variational gradient descent (SVGD) method following~\cite{QL-JL-MJ:16}~and \cite{QL:17}. SVGD is a deterministic sampling algorithm that iteratively transports a set of samples (particles) to approximate a given distribution $p(x)$ on an open set $X\subset\mathbb{R}^d$, based on a gradient-based update that guarantees optimally decreasing the KL divergence within a function space~\cite{QL:17}. Note that we will be using samples and particles interchangeably. To achieve this, we initialize the samples with some simple distribution $q(x)$, and update them via map
\begin{equation}\label{eq::transform}T (x) = x + \epsilon \,\phi(x),\end{equation}
where $\epsilon$ is a small step size, and $\phi(x)$ is a perturbation direction, or velocity field, which should be chosen to maximally decrease the KL divergence of the sample distribution with the target distribution $p(x)$; this is framed by~\cite{QL-JL-MJ:16} as solving a functional optimization problem. Given some structural properties of this optimization problem, which allows drawing connection to ideas in the Stein’s method~\cite{CS:86} used for proving limit theorems or probabilistic bounds in theoretical
statistics, Liu and Wang~\cite{QL-JL-MJ:16} showed that the functional optimization problem that gives us $\phi(x)$ reduces to 
\begin{align}\label{eq::SVGD_orig}
    \mathbb{D}(q\|p):=\max_{\phi\in\mathcal{H}}\{\mathbb{E}_q[\mathcal{S}_p\phi]~~\text{s.t.}~~\|\phi\|_{\mathcal{H}}\leq 1\}
\end{align}
where $\mathcal{H}$ is a normed function space chosen to optimize over. Here, $\mathbb{D}(q\|p)$ is called Stein discrepancy, which provides a discrepancy measure between $q$ and
$p$, with the proven properties that $\mathbb{D}(q\|p)=0$ if $p=q$ and $\mathbb{D}(q\|p)>0$ if $q\neq p$ given $\mathcal{H}$ is sufficiently large.
Furthermore, 
\begin{equation}\label{eq::stein_operator}\mathcal{S}_p\phi(x):=\nabla \log p(x)^\top \phi(x)+\nabla \cdot \phi(x),\end{equation} where $\nabla \cdot \phi(x)=\sum_{k=1}^d \partial_k\phi_k(x)$, and $\mathcal{S}_p$ is a linear operator that maps a vector-valued function $\phi(x)$ to a scalar-valued function $\mathcal{S}_p\phi(x)$, and $\mathcal{S}_p$ is called the Stein operator in connection with the so-called Stein’s identity, which shows that $\mathbb{E}_q[\mathcal{S}_p\phi]=0$ if $q = p$.
\begin{algorithm}

\caption{Stein Variational Gradient Descent}
\label{alg:svgd_orig}
\textbf{Input:} The score function $\nabla_x \log p(x)$.

\textbf{Goal:} A set of particles $\{x_i\}_{i=1}^n$ that approximates $p(x)$.

\textbf{Initialize} a set of particles $\{x_i^{(0)}\}_{i=1}^n$; choose a positive definite kernel $k(x, x')$ and step-size.\\
\textbf{For} iteration \textit{l} \textbf{do}
\begin{align*}
    x_i^{(\ell+1)} \leftarrow x_i^{(\ell)} + \phi^\star(x_i^{(\ell)}) \quad \forall \, i \in\{ 1, \ldots, n\} 
\end{align*}
where,
\begin{align}
\label{eq::svgd_iterate}
\phi^\star(x) = \frac{1}{n}\! \sum\nolimits_{j=1}^n \! \left[ \nabla \log p(x_j) k(x_j, x) \!+\! \nabla_{x_j} k(x_j, x) \right].
\end{align}
\end{algorithm}
The optimization problem~\eqref{eq::SVGD_orig} is an infinite-dimensional functional optimization, which searches for the transformation function $\phi$ in~\eqref{eq::transform} that maps samples drawn from $q$ to $p$. It is critical to select a space $\mathcal{H}$ that is both sufficiently rich and also ensures computational tractability in practice. Literature has shown that the Kernelized Stein discrepancy (KSD) which selects a Reproducing kernel Hilbert space (RKHS) to be $\mathcal{H}$, not only meets this expectation but also goes one step further by yielding a \emph{closed-form} solution~\cite{QL-JL-MJ:16}.



To demonstrate this closed-form solution, let $\mathcal{H}_0$ be a RKHS of scalar-valued functions with a positive definite kernel $k(x, x')$, and $\mathcal{H} = \mathcal{H}_0\times\cdots\times \mathcal{H}_0$ the corresponding $d \times 1$ vector-valued RKHS. Then it can be shown that the optimal solution of~\eqref{eq::SVGD_orig}, denoted by $\phi^\star_{q,p}$, is
\begin{align}\label{eq::opt_phi}
\phi^\star_{q,p}(.) \propto \mathbb{E}_{x\sim q}[\nabla \log p(x)\, k(x,.)+\nabla_x k(x,.)]
\end{align} 
The key advantage of KSD
is its computational tractability: it can be empirically evaluated with samples drawn from $q$ and the gradient $\nabla \log p$, which is independent of the normalization constant in $p$, see~\cite{QL-JL-MJ:16}. In SVGD algorithm, we iteratively update a set of samples using the optimal transform just derived,
starting from a certain initialization. Let $\{x^i_{\ell}\}^n_{i=1}$ be the samples at the $\ell$-th iteration. In this case, the
exact distributions of $\{x^i_{\ell}\}^n_{i=1}$ are unknown or difficult to keep track of, but can be best approximated
by their empirical measure $\hat{q}_{\ell}^n(\text{d}x)=\sum_{i}\delta(x-x_{\ell}^i)\text{d}x/n$. Therefore, it is natural to think that
$\phi^\star_{\hat{q}^n_\ell,p}$, with $q$ in~\eqref{eq::opt_phi} replaced by $\hat{q}^n_{\ell}$, provides the best update direction to move the samples (and 
equivalently $\hat{q}^n_{\ell}$ ) ``closer to" $p(x)$. Implementing this update~\eqref{eq::svgd_iterate} iteratively, we get the main SVGD algorithm in Algorithm~\ref{alg:svgd_orig}.

\section{Stein Coverage: a variational inference approach to multisensor deployment}
\label{sec::propsed_method}

To deploy the robot team $\mathcal{A}$, our goal is to identify representative samples $n\geq N$, called points of interest (PoIs), from the spatial distribution of events $p(x)$. Once the PoIs are identified, we propose a mechanism that assigns PoIs to the sensors in accordance with their heterogeneous spatial coverage ability. 
We call our proposed deployment algorithm {\sf{Stein Coverage}}.

\begin{figure}[t]
    \centering
     \includegraphics[trim=0pt 35pt 0pt 0pt, clip,width=0.5\textwidth]{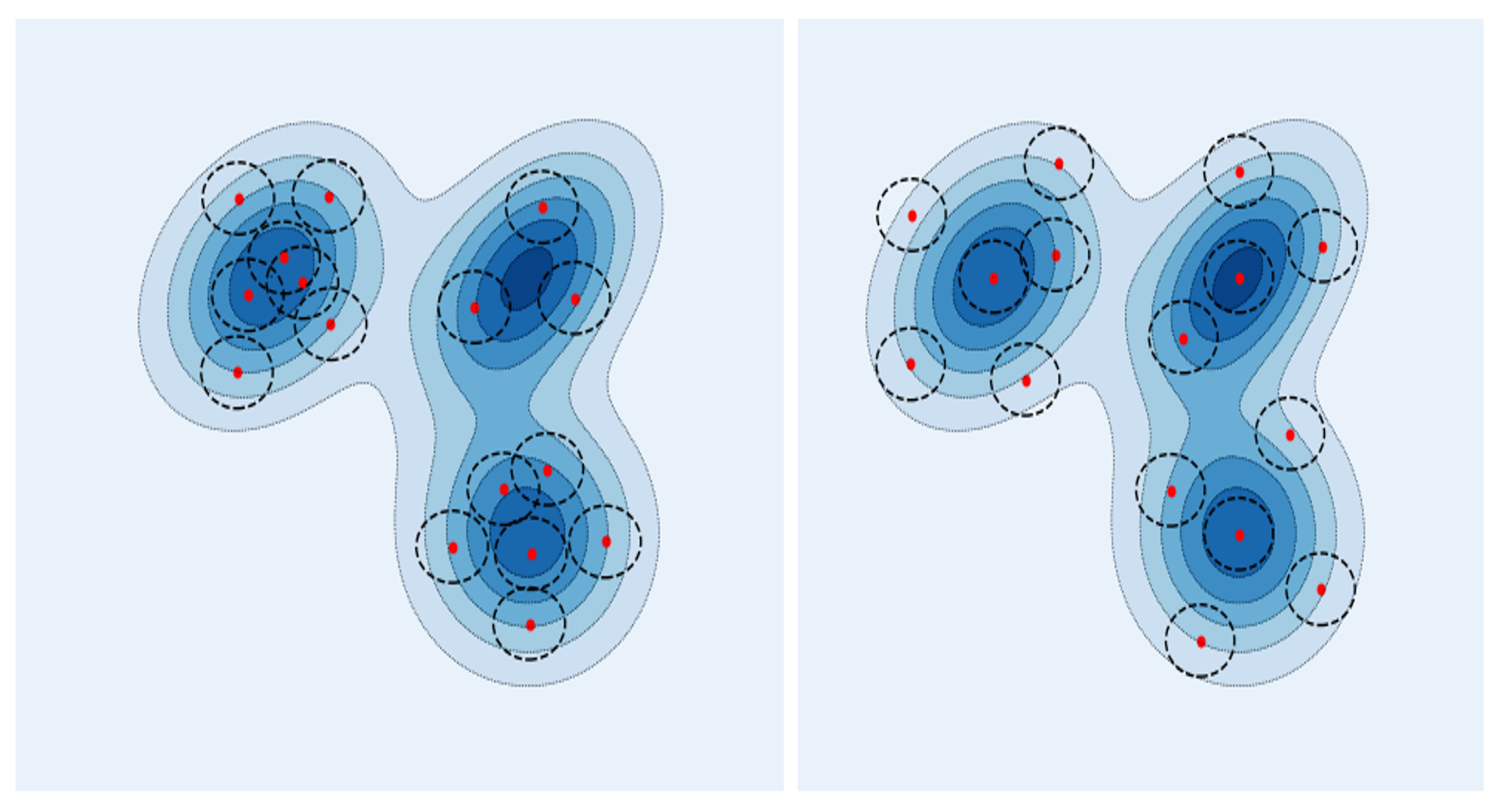}

    \caption{{\small The figure on the left displays the traditional SVGD approach, where the red dots represent the samples placed on a given Gaussian Mixture Model. The dotted circles represent the sensor footprints that overlap when the sensors are installed at the sample points drawn by SVGD. The figure on the right, however, shows the sample placement using our proposed spatially regulated SVGD algorithm, which reduces the overlap of the sensor footprints when placed on the sample points.}}
    \label{fig:stein_vari_algo}
\end{figure}
\subsection{Designing PoIs for robot deployment}
Let $\mathcal{S}=\{1,\cdots,n\}\subset\mathbb{N}$ represent the label set of PoIs, and $\mathbf{x}:=(x_{1}, x_{2}, \cdots,x_{N}) \subset \mathcal{W}^n$ be the location of PoIs. 
The SVGD algorithm can be used to draw these PoIs, but the naive implementation of the SVGD is unaware of the coverage footprint of the sensors, as illustrated in Fig.~\ref{fig:stein_vari_algo}. This can lead to overlapping coverage, which is an inefficient use of the limited coverage resources. 

Given the effective coverage footprint of the sensors, under Assumption~\ref{assump::cover_R}, if any two samples are placed in a way that 
\begin{align}\label{eq::particle_spread}
    g = \mathbb{E}[\Lnorm x_i - x_j\Rnorm^2] - \mathsf{R}^{2} \geq 0 , \quad i \neq j \in \mathcal{N},
\end{align}
the overlap between the coverage of any sensors deployed at $x_i$ and $x_j$ is negligible. We will impose constraint~\eqref{eq::particle_spread}, by informed design of the kernel function $k(x,x')$ of the SVGD Algorithm~\eqref{alg:svgd_orig}.

Intuitively, the update in~\eqref{eq::svgd_iterate} pushes the samples (particles) toward the high probability regions of the event probability via the gradient term $\nabla \log p$, while maintaining a degree of diversity via the second term $\nabla_{x_j} k(x_j,x)$ to prevent the particles from collapsing together in the local Maximum a posteriori (MAP) solutions ~\cite{QL-WD:16}.
In other words, $\nabla_{x_j} k(x_j,x)$  acts as 
a deterministic repulsive force~\cite{QL-WD:16, GD-TC-YS:18}. For our application, we want to choose a kernel that can capture the underlying geometry of the target distribution while also satisfying the pairwise distance constraint~\eqref{eq::particle_spread}. 

For a kernel to be a viable choice in the Stein Variational Gradient Descent formulation, $k(x,.)$ should be in the Stein class of $p$ satisfying 
\begin{equation}\label{eq::kernel_condition}\lim_{\Lnorm x \Rnorm \rightarrow \infty} p(x)\,k(x,.) = 0.\end{equation} Light-tailed distributions satisfy this condition, i.e. smooth functions $k(x,.)$ with proper boundary conditions and bounded target distribution $p(x)$ \cite{QL-JL-MJ:16}. The kernel function that we choose and that satisfies~\eqref{eq::kernel_condition}, is the Mahalanobis-Gaussian kernel function.
\begin{align}\label{Mah_kernel}
    k(x,x') = \exp(- \frac{1}{h} ( x_{i}-x_{j})^{\top} \, W^{-1} \, (x_{i}-x_{j})),
\end{align}
where we want a positive definite matrix $W$ to scale the pairwise distance inside a smooth kernel function. 

We formulate the weighting matrix $W$ 
as a variance induced by the pairwise distance constraint~\eqref{eq::particle_spread}. Notice that 
\begin{align*}
    \sum\nolimits_{i,j}\Lnorm x_{i} - x_{j} \Rnorm^{2} = 2\,n \sum\nolimits_{i=1}^{n}\Lnorm x_{i} - \mu \Rnorm^{2}.
\end{align*}
Now, if the samples are iid, we obtain
\begin{align}
    \label{eq:pdist_var}
    \mathbb{E}(\Lnorm x_{i} - x_{j} \Rnorm^{2}) = 2\,\text{Var}(X_\ell).
\end{align}
where $\mathbb{E}(\Lnorm x_{i} - x_{j} \Rnorm^{2})$ can be represented as the empirical sum of $n^2$ pairwise differences and the empirical variance of the samples is calculated using $\text{Var}(X_\ell) = \frac{1}{n} \sum\nolimits_{i=1}^{n}\Lnorm x_{i} - \mu \Rnorm^{2}  $. Here, $X_\ell$ represents the samples of the empirical distribution approximating the target distribution using the update~\eqref{eq::transform}, and $\mu$ is the mean of these samples. In our formulation, we want the empirical variance that is proportional to the pairwise difference to converge to $\text{Var}(X_\ell) \to \text{Var}(X) \geq \mathsf{R}^2 I$ after $\ell$ iterations. $\text{Var}(X)$ is a $d \times d$ dimensional matrix. Since~\eqref{eq::svgd_iterate}, we are performing the functional gradient descent in RKHS, we need the weighted matrix, $W$ to be represented in the RKHS. Thus, we map the empirical variance using the radial basis function to the RKHS, and define $W = \exp (-\frac{1}{h}\text{Var(X)})$.

\begin{figure}[t]
    \centering
     \includegraphics[width=0.5\textwidth]{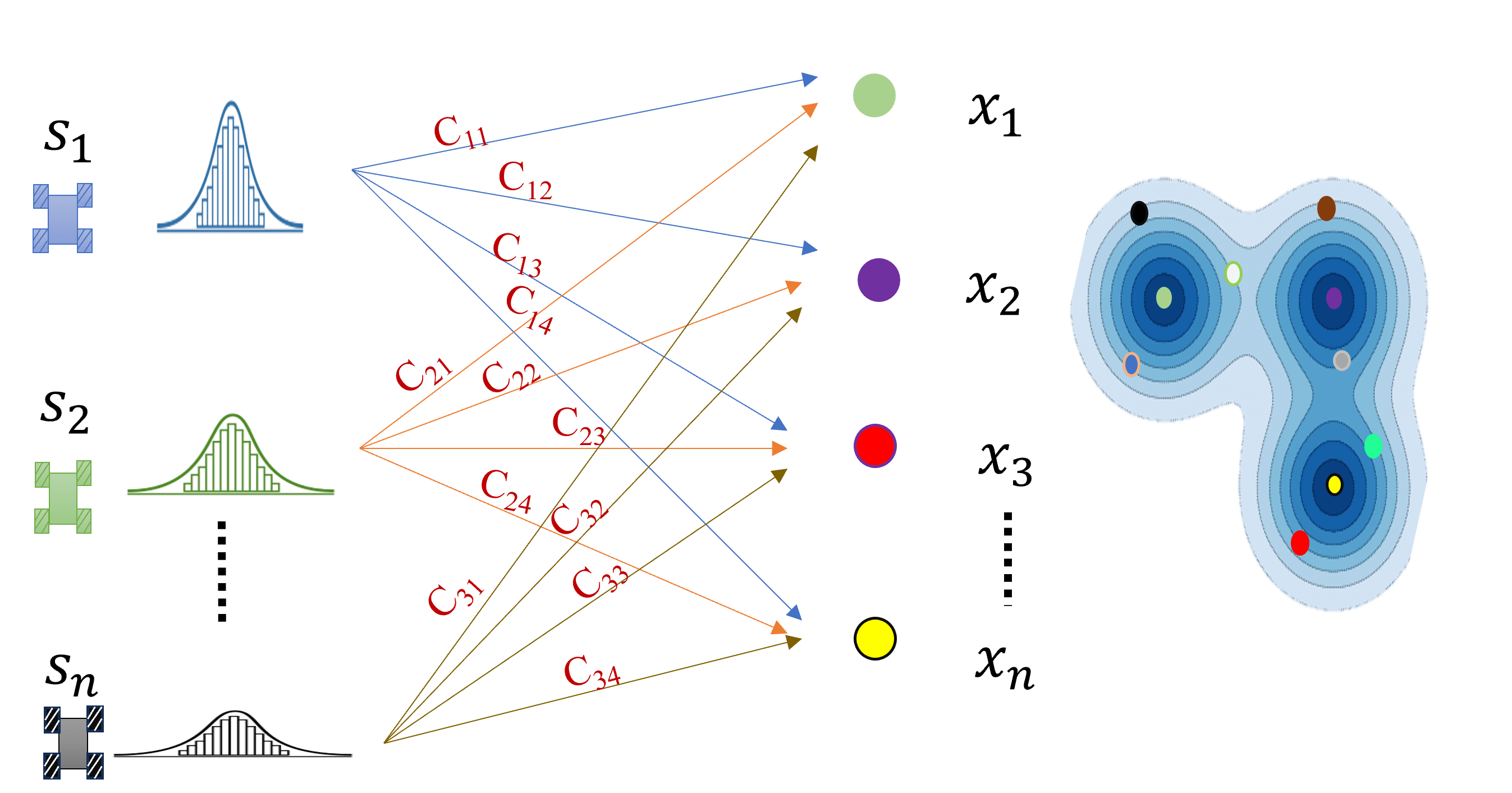}

    \caption{{\small The assignment problem, where each sensor $i \in \mathcal{A}$ with spatial coverage distribution $s_i(x)$ finds the best placement position ${x}_i$ using the cost matrix $C_{ij}$.}}
    \label{fig:assginment} 
\end{figure}
\begin{rem}[Algorithm initialization]\label{rem::initialization}{\rm
Since we want to diversify the PoIs around the event distribution such that there is no overlap, we use our kernel,~\eqref{Mah_kernel} to generate a repulsive force that repels the samples to satisfy the no-overlap condition. However, in doing so we may not have samples in the local modes (MAP), which represent the high region of probability. Thus, we select a few particles in our initialization for which the $\nabla_{x}k(x_{i}^{l},x)$ vanishes and reduces to a typical gradient ascent for MAP. This can be done when we let the parameter $h \rightarrow 0$ for~\eqref{Mah_kernel}, the term $\nabla_{x}k(x_{i}^{l},x)$ from~\eqref{eq::svgd_iterate} vanishes.   
}
\boxend
\end{rem} 
\emph{Selecting PoIs to deploy the robots}: In summary, to select the PoIs to deploy the robots, we will use the SVGD Algorithm~\ref{alg:svgd_orig} with the kernel function~\eqref{Mah_kernel} we designed. 
Our algorithm's initialization is guided by Remark~\ref{rem::initialization}. The selected PoIs are denoted by $\mathcal{X}^\star=\{x_1^\star,\cdots,x_n^\star\}.$

\subsection{Assigning the PoIs to the robots}
For isotropic homogeneous sensors, the number of samples/PoIs extracted by the SVGD process can be set to be the same as the number of sensors ($n=N$). This means that the samples/PoIs will be the final deployment locations for the sensors, and sensors can be assigned to these points randomly.

However, when sensors are heterogeneous and/or have an isotropic or anisotropic quality of service distribution $s(x|x_i,\theta_i)$, $i\in\mathcal{A}$, we need a mechanism to assign sensors to the PoIs and decide the deployment orientation. In what follows, we assume that we search over a finite number of deployment orientations for each sensor, that is, $\theta_i\in\Theta=\{\bar{\theta}_1,\cdots,\bar{\theta}_M\}\subset[0,2\pi]$. We cast our assignment problem as a bipartite matching problem; see Fig.~\ref{fig:assginment}. For every sensor $i\in\mathcal{A}$ and every PoI $j\in\{1,\cdots,M\}$, let 
\begin{align*}
C_{ij}^\star=\min_{\theta_i\in\Theta} \{\mathcal{K}\mathcal{L}(s(x|x_j,\theta_i)\|p(x)) ~~\text{for}~~ x\in\mathcal{C}_i(x|x_i,\theta_i)\},
\end{align*}
which gives us the minimum dissimilarity among all possible orientations $\Theta$ between the service distribution of sensor $i$ and $p(x)$ over the effective coverage footprint of the sensor at the given PoI $j$. In what follows, we let $\theta_i^{j\star}$ be the orientation of sensor $i$ corresponding to $C_{ij}^\star$.
$C_{ij}^\star$ and corresponding $\theta_i^{j\star}$ can also be computed using the discrete sampling method proposed in our previous paper~\cite{DG-KS:23}. With $C_{ij}^\star$'s at hand, we cast the assignment problem as 
\begin{subequations}
\label{eq::assignment}
\begin{align}
&\vect{Z}^\star= \arg\min\sum\nolimits_{j\in\mathcal{S}}\!\sum\nolimits_{i\in\mathcal{A}}\!\!\!Z_{i,j}C_{i,j}^\star,\\
    & Z_{i,j}\in\{0,1\},\quad i\in\mathcal{A},~j\in\mathcal{S},\\
     & \sum\nolimits_{i\in\mathcal{A}}\!\!\!Z_{i,j}=1,~ \forall i\in \mathcal{A},\label{eq::assignment_sensor} \\ &\sum\nolimits_{j\in\mathcal{S}}\!\!\!Z_{i,j}\leq 1, ~ \forall j\in\mathcal{S},\label{eq::assignment_cluster}.
\end{align}
\end{subequations}

The optimization problem~\eqref{eq::assignment} is in the form of a standard assignment problem and can be solved using existing algorithms, such as the Hungarian algorithm~\cite{KW-55}. The solution of the assignment problem gives the final configuration of the sensors $(x_i,\theta_i^\star)$.


\section{Simulation and Experimental Analysis}

\label{sec::num}
This section provides a set of experiments and simulations to study the performance of the proposed framework for homogeneous and heterogeneous sensors with both isotropic and anisotropic configurations. The framework provides a final configuration of the sensor that maximizes the spatial coverage of the environment. To illustrate the collective behavior, we first consider a $50 \times 50$ square environment. The event distribution function $p: [0,50]^{2} \rightarrow \mathbb{R}_{> 0}$ is given by a GMM described by
\begin{align}
    \label{eq::density_function}
    p(x|\psi) \quad  = \quad  \sum\nolimits^{K}_{k=1}\pi_{k} \mathcal{N}(x|\mu_{K}, \Sigma_{k}).
\end{align}
where $K\in\real_{>1}$ is the number of basis of the GMM and  $\psi = ( \pi_{k},\mu_{K}, \Sigma_{k})$ are the mixing weights, mean, and covariance, respectively, for $k\in\{1,\cdots,K\}$. 
\begin{figure}[t]
    \centering
     \includegraphics[width=0.48\textwidth]{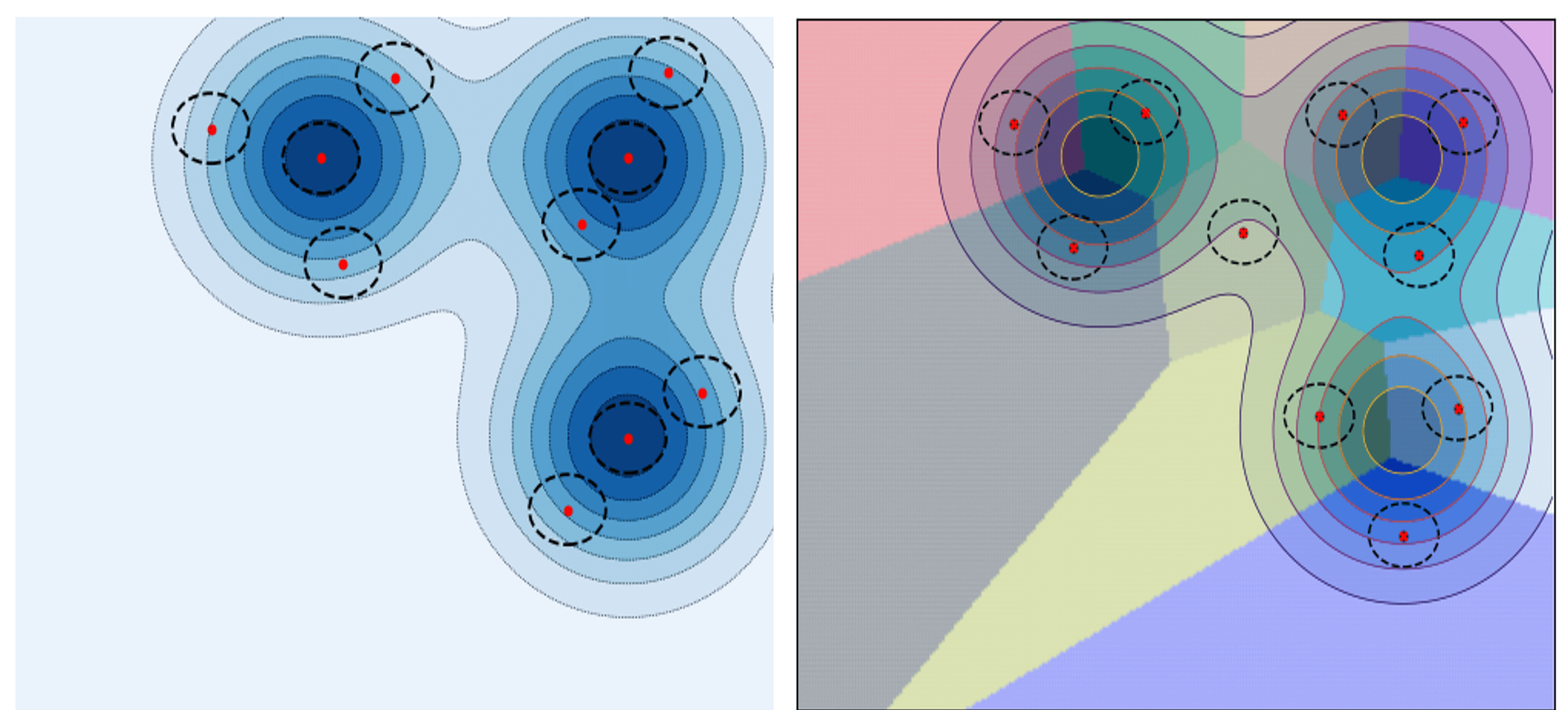} \\
       \includegraphics[width=0.48\textwidth]{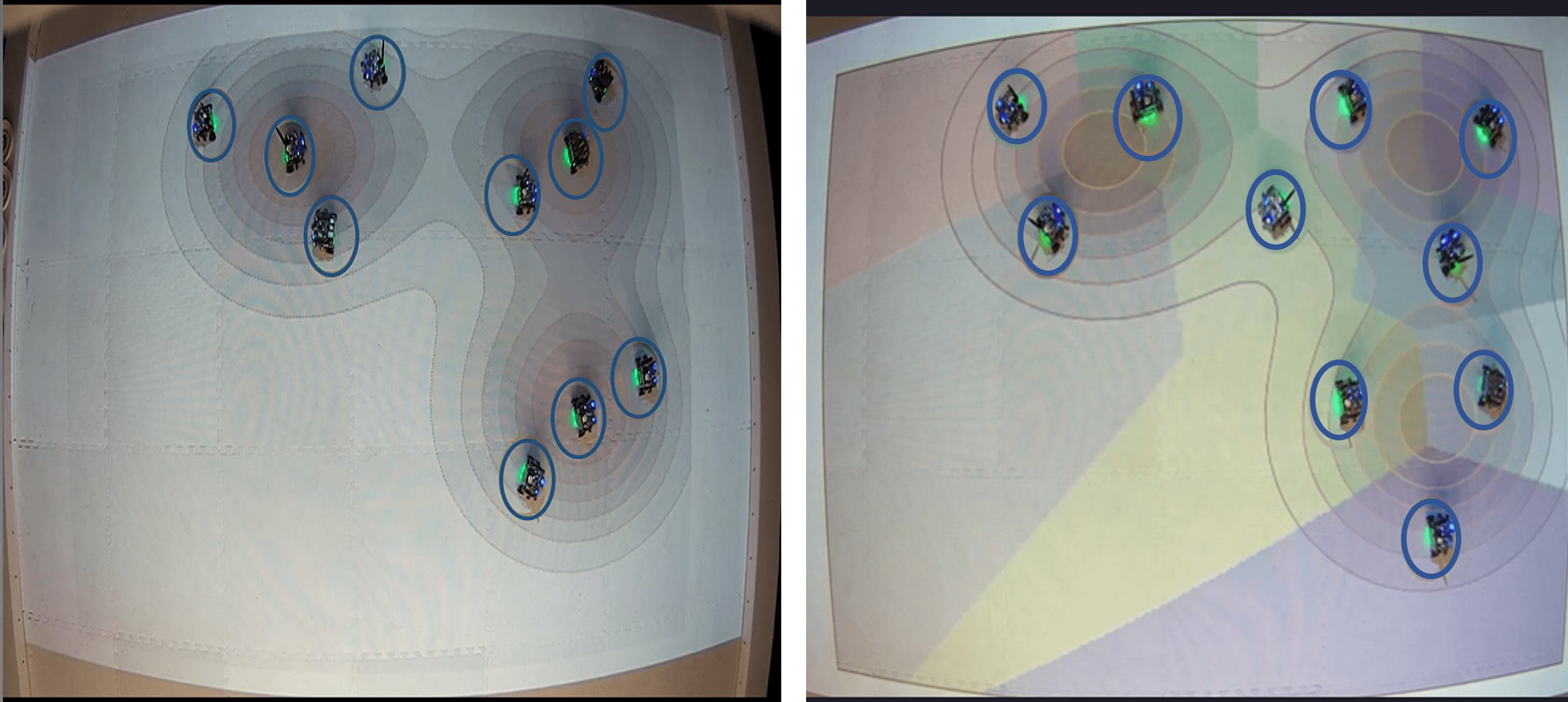} 

    \caption{{\small Sensor deployment for homogeneous sensors: (left) proposed {\sf{Stein Coverage}} method, (right) Voronoi partitioning.
    The plots in the bottom row show the experimental deployment at the Robotarium~\cite{SW-ME:20}, a remotely accessible multi-agent research testbed.}}
    \label{fig:simulation_homogenous} 
\end{figure}

\subsection{Homogeneous isotropic sensors}
First, we consider a group of $N =10$ isotropic homogeneous sensors with a probabilistic sensing model of $s_i(x|x_i,\theta_i)=\mathcal{N}(x|\mu_i,\Sigma_i)$ where $\mu_i=x_i$ and $\Sigma_i=\sigma^2 I$, $\sigma=2.5$. Here, the orientation of the robot, $\theta_i$, does not play a role as the coverage is isotropic. We let the radius of the effective sensing footprint of each sensor be $2\sigma=5$, i.e., $\mathcal{C}_i(x|x_i,\theta_i)=\{x\in\mathcal{W}\,|\,\|x-x_i\|^2\leq 5\}$, which results in $\mathsf{R}=5$. We use our proposed {\sf{Stein Coverage}} algorithm to determine the location of the sensors to cover the event density as shown on the left side of Fig.~\ref{fig:simulation_homogenous}. 
To compare our proposed method with conventional Voronoi partitioning, we obtain the desired Voronoi partitions $ \mathcal{V}(.)= \{\mathcal{V}_{1}, \cdots, \mathcal{V}_{n} \} $, from minimizing  
\begin{align*}
  \sum\nolimits^{n}_{i=1} \int_{V_{i}(x)} \Lnorm x - x_{i} \Rnorm ^{2} p(x) dx .
\end{align*}
where $p(x)$ is given in~\eqref{eq::density_function}, and is used to bias the importance of the location optimization function. The result is shown in the plots on the right-hand side of Fig.~\ref{fig:simulation_homogenous}. 
Because the sensors are homogeneous, there is no need to solve the optimal assignment~\eqref{eq::assignment}. Comparing the two configurations shows us that {\sf{Stein Coverage}} is able to capture the high-density region of the density function $\phi$, compared to Voronoi partitioning. This is because SVGD uses a kernelized (functional) gradient descent method, to move the particles where the KL divergence is minimized in terms of Stein discrepancy, the update is better at capturing the mean and variance of Gaussian distributions. However, Voronoi partitioning based on Euclidean distance might not capture the underlying structure of the distribution well. 

\begin{figure}[t]
    \centering
    \begin{subfigure}[b]{0.45\textwidth}
        \includegraphics[width=\textwidth]{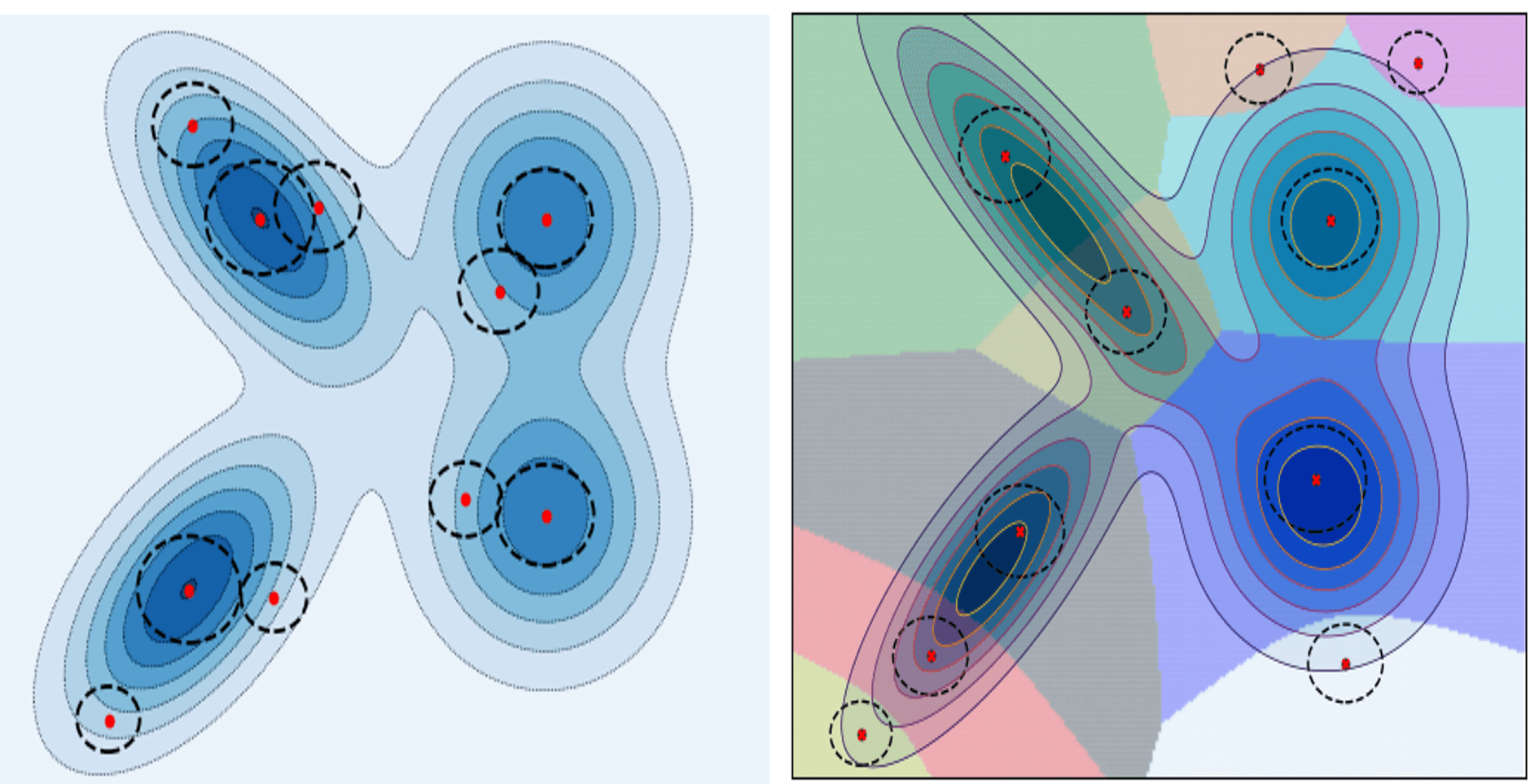}
        \caption{{\small Sensor deployment for heterogeneous sensors: (left) {\sf{Stein Coverage}}, (right) Voronoi partitioning using power diagrams.}}
        \label{fig:simulation_heterogenous}
    \end{subfigure}
    \hfill
    \begin{subfigure}[b]{0.45\textwidth}
        \includegraphics[width=\textwidth]{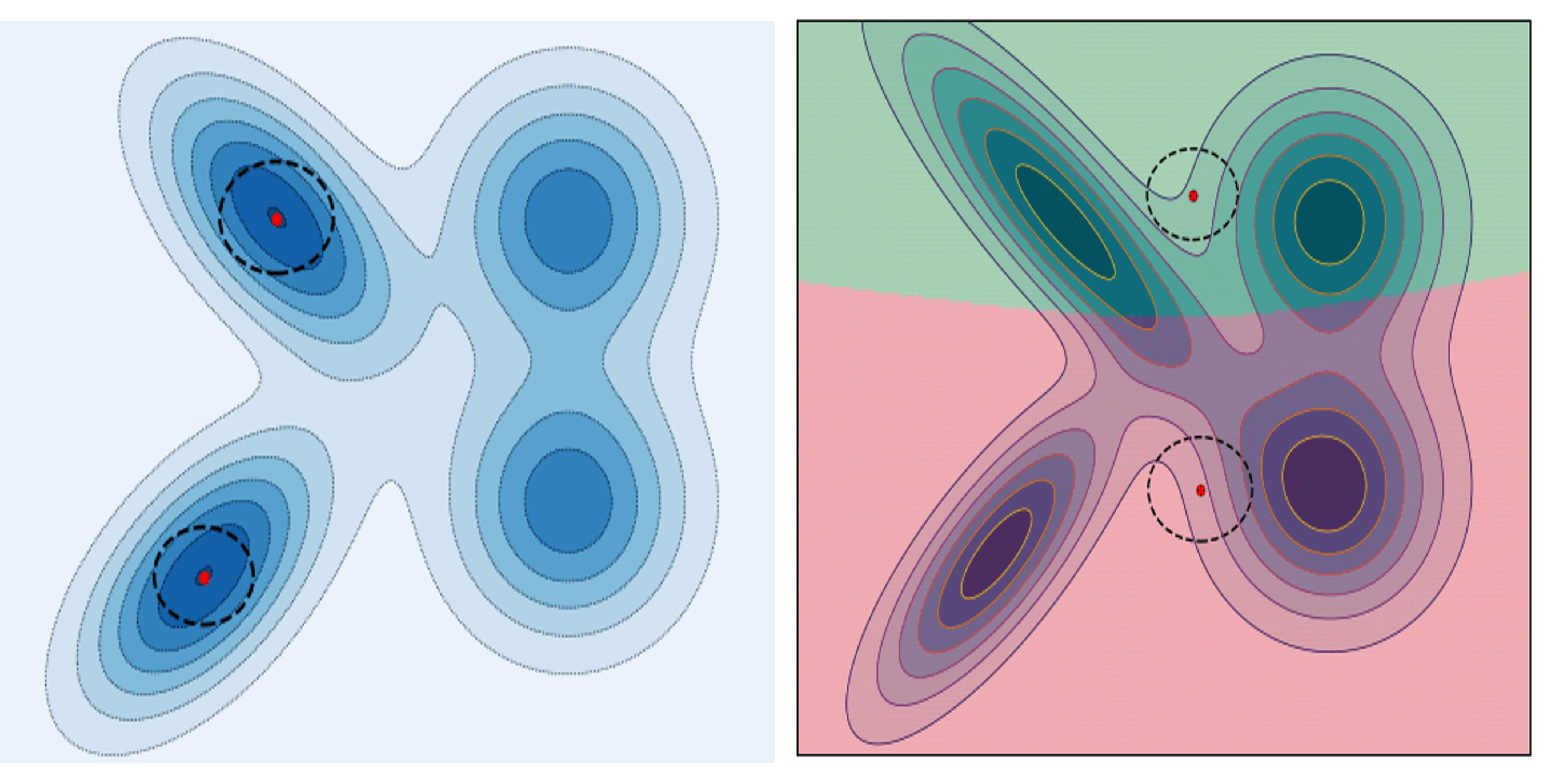}
        \caption{{\small Comparison of sensor deployment:(left) proposed {\sf{Stein Coverage}}, (right) Voronoi partitioning using power diagrams, when the number of sensors is less than the number of clusters.}}
        \label{fig:simulation_heterogenous_2ag}
    \end{subfigure}
    \caption{{\small Sensor deployment for heterogeneous sensor.}}
    \label{fig:sensor_deployments}
\end{figure}

\subsection{Heterogeneous isotropic and anisotropic sensors}
Next, we present a simulation for a group of $N=10$ heterogeneous sensors. Here, we consider a scenario where the sensors are isotropic with uniform coverage everywhere, so we can compare our results with the power diagrams method, a generalized Voronoi partitioning deployment for heterogeneous sensors with uniform quality of service distribution. 

\begin{figure}[t]
    \centering
    \begin{subfigure}[b]{0.45\textwidth}
        \includegraphics[width=\textwidth,height=1.75in]{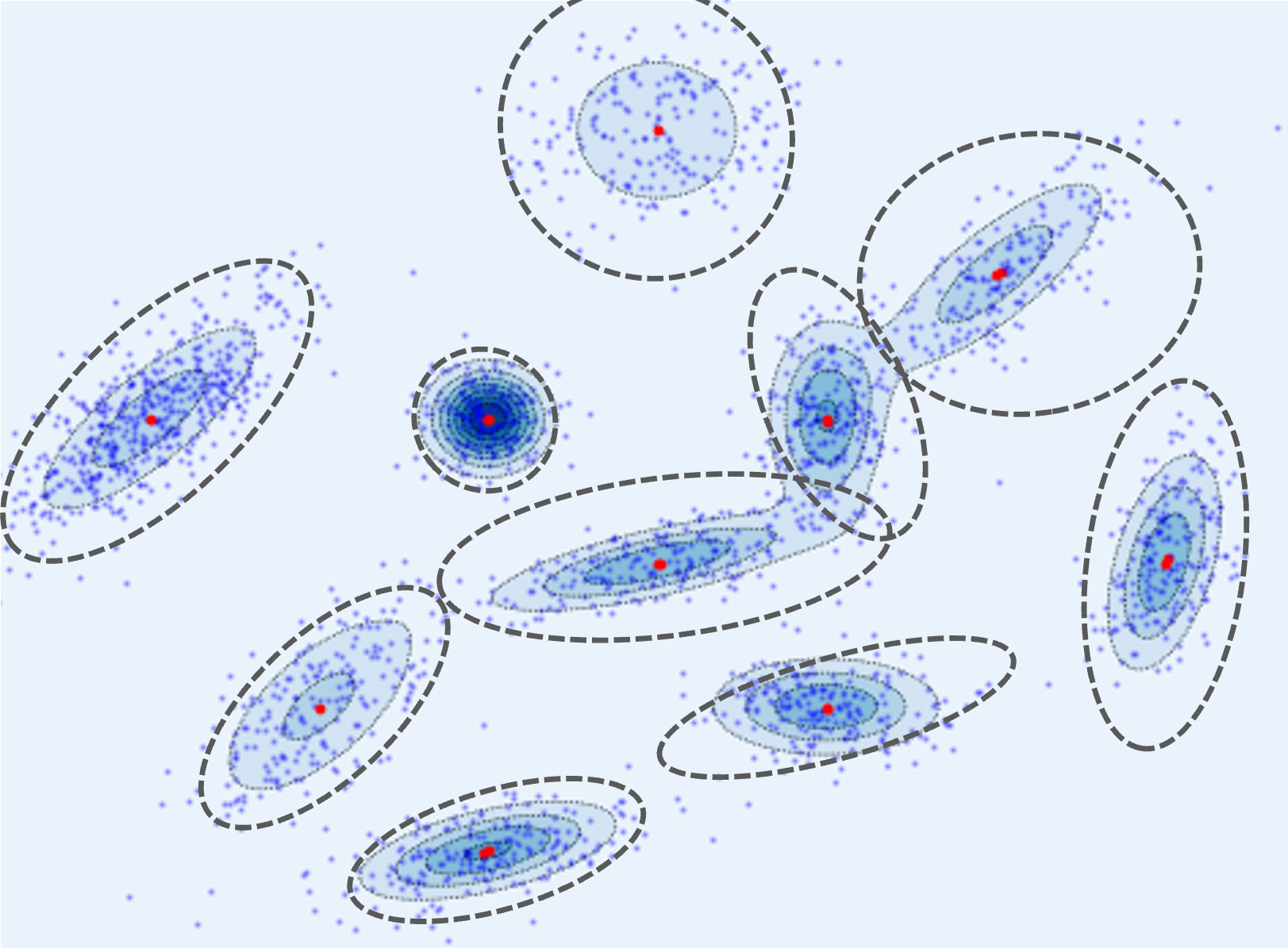}
        \caption{{\small Sensor deployment for anisotropic heterogeneous sensors. The level sets represent the multimodal distribution of the target points (blue dots) and the ellipsoid encircling the level sets represents the configuration of anisotropic footprint calculated using the proposed algorithm.}}
        \label{fig:anisotropic_a}
    \end{subfigure}
    \hfill
    \begin{subfigure}[b]{0.45\textwidth}
        \includegraphics[width=\textwidth,height=1.75in]{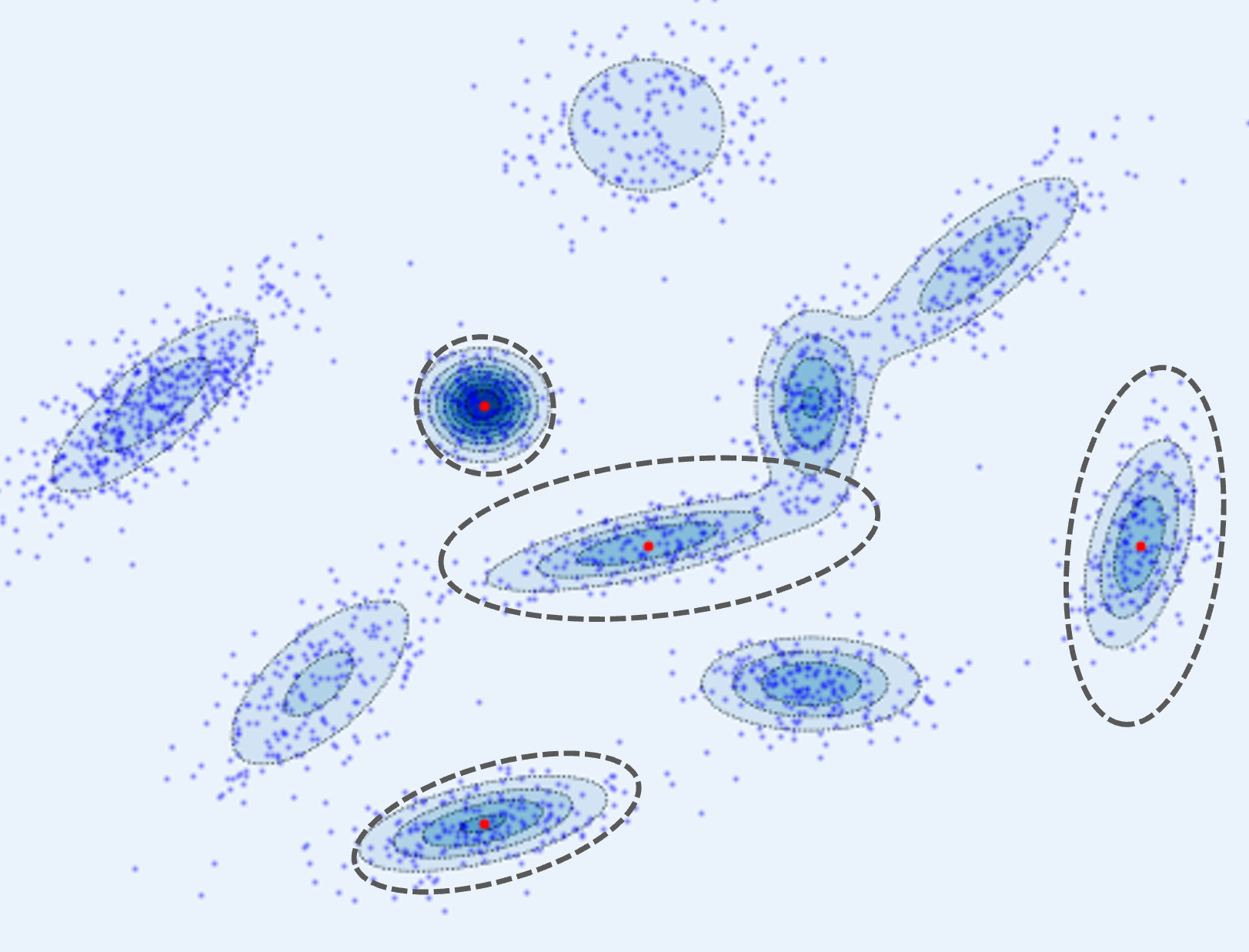}
        \caption{{\small Sensor deployment for heterogeneous anisotropic sensors when the number of sensors is less than the number of clusters.}}
        \label{fig:anisotropic_4ag}
    \end{subfigure}
    \caption{{\small Heterogeneous and anisotropic sensor deployment}}
    \label{fig:anisotropic_deployments}
\end{figure}
We assume that every sensor's effective footprint is $\mathcal{C}_i(x|x_i,\theta_i)=\{x\in\mathcal{W}|\|x-x_i\|\leq r_i\}$ for $i\in\mathcal{A}$ with corresponding $r_i$ belonging to the randomly generated ordered set $\{2.2, 2.3, 2.5, 2.8, 2.8, 3.0, 3.3, 3.4, 3.6, 3.8\} \subset\mathbb{R}_{> 0}$. 
We compare this method with coverage for heterogeneous sensors, using power diagrams, $ \mathcal{P}(.)=\{\mathcal{P}_{1},\cdots, \mathcal{P}_{n} \}$ generalized Voronoi partitioning with additive weights$ (r_{i} \in $ $\mathbf{r})$ obtained from minimizing 
\begin{align*}
    \sum\nolimits^{n}_{i=1} \int_{\mathcal{P}_{i}} (\Lnorm x - x_{i}  \Rnorm ^{2} - r_{i}^2) p(x) dx .
\end{align*}
The result of implementing the {\sf{Stein Coverage}} and deployment using the power diagrams is shown in Fig.~\ref{fig:simulation_heterogenous}. As seen, the {\sf{Stein Coverage}} achieves a more effective placement, with sensors deployed to the important regions and in accordance to their coverage footprint. The difference between the two methods is even more evident when the number of sensors is significantly less than important coverage subject clusters; an example scenario is shown in Fig.~\ref{fig:simulation_heterogenous_2ag}. As seen in Fig.~\ref{fig:simulation_heterogenous_2ag}, a Voronoi-based solution is far less effective in this case, because of its implicit consideration of coverage beyond the effective footprint of the sensors (enforcing the footprint as a soft constraint). 



Next, we present a simulation for $N = 10$ heterogeneous sensors, where the footprints are anisotropic with probabilistic sensing models of $s_i(x|x_i,\theta_i)=\mathcal{N}(x|\mu_i,\Sigma_i)$ where $\mu_i=x_i$ and $\vect{\Sigma_i}=\left[\begin{smallmatrix}
\sigma^2_{i,11} & \sigma^2_{i,12}\\
\sigma^2_{i,21} & \sigma^2_{i,22}
\end{smallmatrix}\right]$
,where $\sigma_{i,11}\in\{2,2,2,1,1,1,2,1,2,2\}$, $\sigma_{i,12}\in\{1,0.5,1,2,0,0,0,0,1,0\}$, $\sigma_{i,21}\in\{1,1,0.1,0,1,0,0,0,1,0\}$ and, $\sigma_{i,22}\in\{2,2,1,1,2,1,1,2,2,2\}$. Here, the orientation of the robot, $\theta_i$, is important and should be taken into account when designing the final sensor deployment configuration. We let the radius of the effective sensing footprint of each sensor $i\in\{1,\cdots,N\}$ be $2\max\{\sigma_{i,1},\sigma_{i,1}\}$, and use $\mathsf{R} = 4$.

The density distribution $p(x)$ is a GMM with $K=10$ basis, as shown in Fig.~\ref{fig:anisotropic_a}. In this case, because the sensors are anisotropic, we want to find the configuration $(x_i,\theta_i)$, $i \in \{1,\cdots,N\}$, to maximize the coverage. In implementing {\sf{Stein Coverage}}, we use a sampling-based method proposed in our previous work~\cite{DG-KS:23} to compute $C_{ij}^\star$ and $\theta_i^{j\star}$. Fig.~\ref{fig:anisotropic_a} shows the deployment configuration $(x_i,\theta_i)$ when the number of sensors is equal to the number of basis of the  GMM distribution $p(x)$. On the other hand, Fig.~\ref{fig:anisotropic_4ag}, illustrates the deployment configuration $(x_i,\theta_i)$, when we have a limited number of sensors ($N=4$) compared to the number of the basis ($K=10$). As seen, {\sf{Stein Coverage}} deploys the sensors to most important areas and determines their orientation to provide the best local distribution-matching between the sensors and the coverage subjects distribution.

\section{Conclusion}

In this paper, we consider the spatial coverage optimization problem of multiple sensors and propose a novel framework inspired by Stein Variational Gradient Descent to configure a group of sensors to optimally cover a given event distribution in convex environments. We use a new statistical distance measure to determine potential sensor assignment locations. We characterized a locally optimal coverage configuration with soft workspace statistics and compared it with hard workspace statistics like Voronoi partitioning. We compared these two different approaches using numerical simulations for both heterogeneous and homogeneous sensors. In numerical simulations, we were able to show how the coverage quality of the proposed framework is better compared to the Voronoi partitioning approaches. Finally, we extend our work to deploy the sensor with an anisotropic footprint using the method we proposed in our previous paper. Future work will explore the practical extension of our framework to non-convex environments and 3D coverage for environmental monitoring, surveillance, and search and rescue. 

\bibliographystyle{ieeetr}%

\end{document}